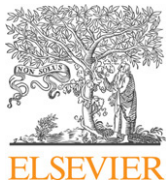
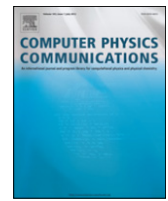

# ElecSus: A program to calculate the electric susceptibility of an atomic ensemble

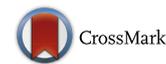

Mark A. Zentile [*], James Keaveney, Lee Weller [1], Daniel J. Whiting, Charles S. Adams, Ifan G. Hughes

*Joint Quantum Center (JQC) Durham-Newcastle, Department of Physics, Durham University, South Road, Durham, DH1 3LE, United Kingdom*

**A B S T R A C T**

We present a computer program and underlying model to calculate the electric susceptibility of a gas, which is essential to predict its absorptive and dispersive properties. Our program focuses on alkali-metal vapours where we use a matrix representation of the atomic Hamiltonian in the completely uncoupled basis in order to calculate transition frequencies and strengths. The program calculates various spectra for a weak-probe laser beam in an atomic medium with an applied axial magnetic field. This allows many optical devices to be designed, such as Faraday rotators/filters, optical isolators and circular polarisation filters. Fitting routines are also provided with the program which allows the user to perform optical metrology by fitting to experimental data.

**Program summary**

*Program title:* ElecSus

*Catalogue identifier:* AEVD_v1_0

*Program summary URL:* http://cpc.cs.qub.ac.uk/summaries/AEVD_v1_0.html

*Program obtainable from:* CPC Program Library, Queen's University, Belfast, N. Ireland

*Licensing provisions:* Apache License, version 2

*No. of lines in distributed program, including test data, etc.:* 191 270

*No. of bytes in distributed program, including test data, etc.:* 309 4994

*Distribution format:* tar.gz

*Programming language:* Python.

*Computer:* Any single computer running Python 2.

*Operating system:* Linux, Mac OSX, Windows.

*RAM:* Depends on the precision required and size of the data set, but typically not larger than 50 MiB.

*Classification:* 2.2, 2.3.

*External routines:* SciPy library [1] 0.12.0 or later, NumPy [1], matplotlib [2]

*Nature of problem:*
Calculating the weak-probe electric susceptibility of an alkali-metal vapour. The electric susceptibility can be used to calculate spectra such as transmission and Stokes parameters. Measurements of experimental parameters can be made by fitting the theory to data.

*Solution method:*
The transition frequencies and wavelengths are calculated using a matrix representation of the Hamiltonian in the completely uncoupled basis. A suite of fitting methods are provided in order to allow user supplied experimental data to be fit to the theory, thereby allowing experimental parameters to be extracted.

*Restrictions:*
Only describes a magnetic field parallel to the laser beam propagation direction. Results are only valid in the weak-probe regime.

*Running time:*
At standard precision less than a second for a theory curve, fitting will take 10 s to 20 min depending on the method used, the number of parameters to fit and the number of data points.

*References:*
[1] T.E. Oliphant, Comput. Sci. Eng. 9, 10 (2007). http://www.scipy.org/.
[2] J.D. Hunter, Comput. Sci. Eng. 9, 10 (2007). http://matplotlib.org/.

## 1. Introduction

Atomic physics of thermal vapours is an expanding field of interest, ranging from the fundamental to the applied. Examples of fundamental physics include observations of the cooperative Lamb shift [1], hyperfine Paschen–Back regime [2], macroscopic entanglement [3], collisional laser cooling [4] and dipole–dipole induced bistability [5] to name a few. Applications include compact and precise magnetometers [6] and clocks [7], laser frequency stabilisation both on [8] and off-resonance [9,10], enhanced frequency up-conversion [11], trans-spectral orbital angular momentum transfer [12] and quantum memories [13,14].

Most applications benefit from being able to predict the absorptive and dispersive properties (in absolute measures) of the medium. The electric susceptibility is key in calculating these properties [15]; here we present a fast, and easy to use, computer program based on the electric susceptibility of an atomic ensemble, that can be used to predict absorption and dispersion given certain parameters. This facilitates designing optical devices, such as Faraday filters [16–20] and optical isolators [21], without resorting to experimental trial and error. Also, since each theory curve will typically take less than a second to compute, fitting experiment to theory becomes practical. This allows experimental parameters to be measured efficiently. This has applications for optical magnetometry [22], optical thermometry [23–25], number density measurements in optically thick vapour [26,27], and diagnostics for devices such as vacuum dispensers [28–30] and vapour cells [31–35].

For the particular case of alkali-metal vapours we have built up a theoretical model of the electric susceptibility [15] that includes dipole–dipole induced linewidth broadening [36] and axial magnetic fields [37]. The model has been shown to be accurate at the ∼0.5% rms level when calculating the frequency dependence of transmission [15,36] and Faraday rotation [10,21,37] for applied magnetic fields of up to ∼6 kG, and is not expected to break down until the magnetic interaction becomes comparable to the fine structure interaction. Fitting to the model has found utility in measuring number density [38] and large magnetic fields with high precision [10], and estimating buffer gas pressures in vapour cells [39]. Being able to extract shifts of many atomic transitions that overlap due to Doppler and self broadening was also key in observing the cooperative Lamb shift [1] and studying atom–surface interaction [40], where a similar model was used.

In principle, similar models could be constructed for any gas, but here we focus on alkali-metal vapours. Extensions to the case of molecular gases are also of interest [41].

## 2. Theoretical background

The complex index of refraction, $n_c \equiv n + i\beta$, of an optical medium is related to its electric susceptibility, $\chi \equiv \chi_{re} + i\chi_{im}$, by the equation [42]

$$n_c = \sqrt{1+\chi} \approx 1 + \frac{\chi_{re}}{2} + i\frac{\chi_{im}}{2}, \qquad (1)$$

where the approximation is valid when $|\chi|$ is small.[2] The real part of $n_c$ (commonly known as the refractive index), $n$, gives the ratio of the speed of light in vacuum, $c$, to the phase velocity in the medium. The attenuation coefficient, $\alpha$, is obtained from the imaginary part, $\beta$, by $\alpha = 2k\beta \approx k\chi_{im}$ [42], where $k$ is the angular wavenumber. The electric susceptibility of a medium is a function of frequency and hence gives rise to dispersion, as well as frequency-dependent attenuation.

The spectral dependence of $n$ is also important in calculating the propagation of optical pulses in the medium. The group index, $n_g$, which is defined as $c$ divided by the group velocity can be calculated if we know $n$ as a function of angular frequency $\omega$ by using the relation [43],

$$n_g = n + \omega \frac{\partial n}{\partial \omega}. \qquad (2)$$

Calculating the group index allows one to estimate the speed of an optical pulse in the medium, and therefore is a useful quantity for slow light [44] or fast-light [45,46] experiments.

Both the real and imaginary parts of the electric susceptibility have characteristic line-shapes which we will look at in the next section.

### 2.1. The electric susceptibility line-shape

Without motion of the atoms, for a single atomic transition labelled $i$, we can write [15],

$$\chi_i(\Delta_i) = \frac{C_i^2 d^2 \mathcal{N}_a}{\epsilon_0 \hbar} f(\Delta_i) \qquad (3)$$

$$f(\Delta_i) = \frac{i}{\Gamma/2 - i\Delta_i}, \qquad (4)$$

where $C_i^2$ is the relative strength factor of the transition, $d$ is the reduced dipole matrix element (see Section 2.2.1), $\epsilon_0$ is the vacuum permittivity, $\Gamma$ is the decay rate from the excited state of the transition, $\hbar$ is the reduced Planck's constant and $\Delta_i$ is the angular detuning from resonance, defined as $\Delta_i \equiv \omega - \omega_i$ where $\omega$ is the angular frequency of the light and $\omega_i$ is the resonance angular frequency of transition $i$. $\mathcal{N}_a$ is the number density of identical atoms and can be calculated from the elemental number density $\mathcal{N}$ (see Appendix A) by the following equation,

$$\mathcal{N}_a = \frac{F_a \mathcal{N}}{2(2I+1)}, \qquad (5)$$

---
[2] When $|\chi|$ is small, local field effects are correspondingly small [42].

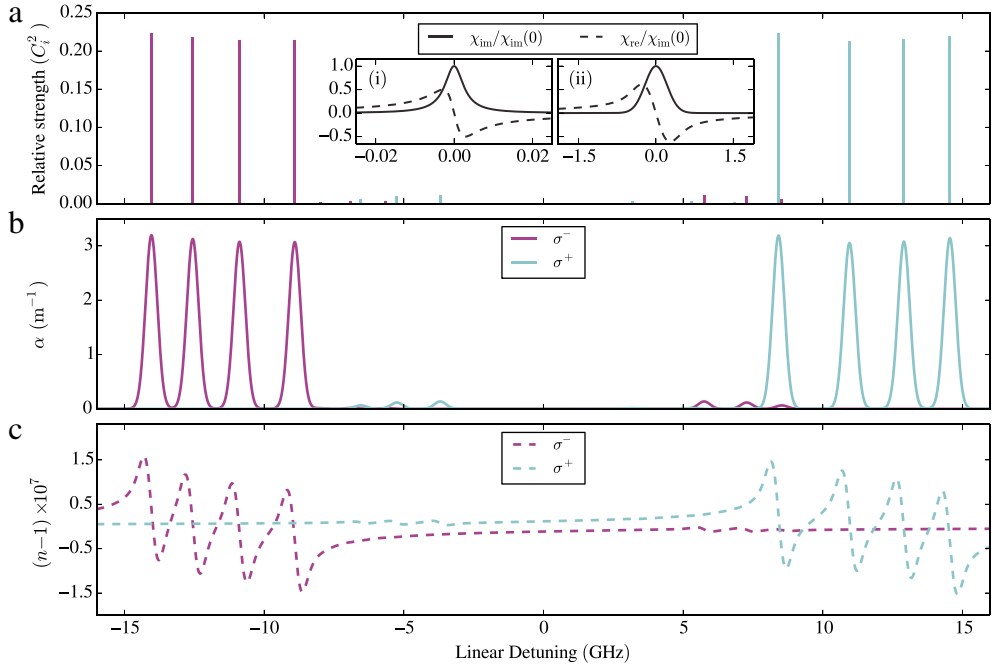

**Fig. 1.** (Colour online) Graphs showing how the refractive indices and absorption coefficients are found as a function of global detuning ($\Delta/2\pi$). Panel (a) is a stick spectrum showing the transition frequencies and strengths which are calculated from finding the eigenenergies and eigenvectors of the Hamiltonian of $^{87}$Rb in a 6 kG magnetic field. Inset (i) of panel (a) shows the Doppler-free complex lineshape profile for $^{87}$Rb on the D$_1$ line (795 nm). Doppler broadening is added by convolving the lineshapes in (i) with a Gaussian resulting in the lineshapes given in (ii), (both insets are normalised to the peak of their respective imaginary part). For this example the Gaussian distribution was calculated for a temperature of 20 °C. The lineshapes in (ii) are then simply added to the spectrum at the position of the transitions multiplied by their strength. This yields the extinction coefficient shown in panel (b) or the refractive index shown in panel (c).

where $F_a$ is the isotopic fraction and $I$ the nuclear spin quantum number. The denominator, $2(2I + 1)$, is the number of states in the ground manifold since we expect the atoms to be distributed evenly among all these states. This is justified by calculating the Boltzmann factor. For large magnetic fields (∼5000 G) and low temperatures for thermal vapours (20 °C), the population of states in the ground manifold deviates from uniform at the 0.1% level. The assumption of zero population in the excited manifolds is again justified by calculating the Boltzmann factor which shows the population is of the order of $10^{-13}$. Note that when not in the weak-probe regime, hyperfine pumping will occur and redistribute populations [47,48].

The real and imaginary parts of $\chi_i$ are plotted as a function of $\Delta_i$ in panel (i) of Fig. 1. The decay rate, $\Gamma$, is also the full width half maximum of the imaginary part of $\chi_i$ (a Lorentzian distribution). $\Gamma$ has contributions from natural broadening, dipole–dipole induced self broadening [36], and any extra homogeneous broadening e.g. pressure broadening due to buffer gases. The centre of the complex line-shape occurs at the resonance frequency, however due to motion, the Doppler effect causes the atom to observe a shifted frequency. A thermal ensemble of atoms has a Gaussian distribution of velocities in the light propagation direction, given by [49]

$$g(v) = \frac{1}{U\sqrt{\pi}} \exp\left(-\frac{v^2}{U^2}\right), \tag{6}$$

where $v$ is the component of velocity of the atom in the light propagation direction and $U$ is the root mean square (rms) speed of the atoms, given by $U = \sqrt{2k_B T/m}$, where $k_B$ is Boltzmann's constant, $T$ is the thermodynamic temperature and $m$ is the mass of the atom. The resulting atomic line-shape is therefore given by a convolution between $f$ and $g$,

$$\chi_i(\Delta_i) = \frac{C_i^2 d^2 \mathcal{N}_a}{\epsilon_0 \hbar} \mathcal{V}(\Delta_i) \tag{7}$$

$$\mathcal{V}(\Delta_i) = \int_{-\infty}^{\infty} f(\Delta - kv)g(v)\mathrm{d}v, \tag{8}$$

Panel (ii) of Fig. 1 shows the line-shape of $\chi_i$ when we include Doppler broadening at 20 °C. It is important for the atom–light interaction to be in the weak-probe regime or deviations from this line-shape will be seen [50,51]. Further deviations can also occur for large amounts of buffer gases or short length cells due to Dicke narrowing [52]. These effects are beyond the scope of this work.

To calculate the total susceptibility we add the contribution from all transitions ($\chi = \sum_i \chi_i$). We then write the susceptibility in terms of a global detuning, $\Delta$, which is the frequency relative to a global linecentre. For convenience we have chosen zero detuning to occur at the weighted linecentre for D$_1$ ($n^2S_{1/2} \rightarrow n^2P_{1/2}$) and D$_2$ ($n^2S_{1/2} \rightarrow n^2P_{3/2}$) transitions (see Appendix B) where $n = 3, 4, 5$ or 6 for sodium, potassium, rubidium or caesium, respectively. We now look at using an atomic Hamiltonian to calculate the frequencies of the transitions relative to the linecentre as well as their strengths.

### 2.2. The atomic Hamiltonian

The atomic Hamiltonian is written as a sum of interaction mechanisms,

$$H = H_0 + H_f + H_{hf} + H_Z, \tag{9}$$

where $H_0$ is the coarse atomic energy and $H_f$, $H_{hf}$ and $H_Z$ are the fine, hyperfine and external magnetic field interactions respectively. The fine structure interaction can be written as

$$H_f = \frac{\gamma_f}{\hbar^2} (\boldsymbol{L} \cdot \boldsymbol{S}), \tag{10}$$

where $\boldsymbol{L}$ and $\boldsymbol{S}$ are the orbital and spin angular momenta of the valence electron respectively, and $\gamma_f$ is the spin–orbit constant for the particular atom. The hyperfine interaction has contributions from the magnetic dipole interaction, $H_d$, and electric quadrupole interaction, $H_q$, ($H_{hf} = H_d + H_q$). We have omitted higher order multipole interactions since their effect is small ($\Delta E/h < 1$ kHz [53]). The magnetic dipole interaction can be written as [54]

$$H_d = \frac{A_{hf}}{\hbar^2}(\mathbf{I} \cdot \mathbf{J}), \tag{11}$$

where $A_{hf}$ is the magnetic dipole constant, and $\mathbf{I}$ and $\mathbf{J}$ are the nuclear spin and total electron ($\mathbf{J} = \mathbf{L} + \mathbf{S}$) angular momenta respectively. Using the convention that the magnitude of an arbitrary angular momentum $\mathbf{K}$ is $\sqrt{K(K+1)}\hbar$, the electric quadrupole interaction can be written as [54]

$$H_q = \frac{B_{hf}}{\hbar^4} \left[ \frac{3(\mathbf{I}\cdot\mathbf{J})^2 + \frac{3}{2}(\mathbf{I}\cdot\mathbf{J})\hbar^2 - I(I+1)J(J+1)\hbar^4}{2I(2I-1)J(2J-1)} \right], \tag{12}$$

where $B_{hf}$ is the electric quadrupole constant. See Appendix B for values for $A_{hf}$ and $B_{hf}$. The interaction with an external magnetic field is given as

$$H_Z = \frac{\mu_B}{\hbar}(g_L \mathbf{L} + g_S \mathbf{S} + g_I' \mathbf{I}) \cdot \mathbf{B}, \tag{13}$$

where $\mu_B$ is the Bohr magneton, and $g_L$, $g_S$ and $g_I'$ are the $g$-factors corresponding to the electron orbital, electron spin and nuclear angular magnetic moments. $g_L$ is taken to be 1; the values for the other $g$-factors are given in Appendix B. If we choose our quantisation axis to be parallel to the magnetic field, Eq. (13) reduces to

$$H_Z = \frac{\mu_B}{\hbar}(g_L L_z + g_S S_z + g_I' I_z) B_z. \tag{14}$$

Once the Hamiltonian is constructed, we need to find the eigenenergies, $E_j$, and corresponding eigenstates, $|j\rangle$, in order to calculate the transition energies and strengths. The transition energy is simply the difference in energy between the ground and excited states ($E_e - E_g$).

#### 2.2.1. Calculating transition strengths from the eigenstates

The eigenstates will be given by some combination of completely uncoupled basis states $|L, m_L, m_S, m_I\rangle$, where $m_L$, $m_S$ and $m_I$ are the $z$-projection quantum numbers. Transition strengths are given by the electric dipole matrix element squared, $|\langle g|er_q|e\rangle|^2$, where $er_q$ is the component of the dipole operator in the spherical basis (see Eq. 5.17 in [55]); this chooses whether the transition is $\sigma^+$, $\sigma^-$ or $\pi$ ($\Delta m_L = -q$ where $q = -1, +1$ or 0 respectively). To illustrate how the strength is calculated, we take the example of a $\sigma^-$ transition between the eigenstates

$$|g\rangle = a_1|0, 0, +1/2, -1/2\rangle + a_2|0, 0, -1/2, +1/2\rangle \tag{15}$$

$$|e\rangle = b_1|1, -1, +1/2, -1/2\rangle + b_2|1, +1, -1/2, -3/2\rangle$$
$$+ b_3|1, 0, -1/2, -1/2\rangle + b_4|1, 0, +1/2, -3/2\rangle \tag{16}$$

where $a_1, a_2, b_1, b_2, b_3$ and $b_4$ are parameters which are known after finding the eigenenergies and eigenstates of the Hamiltonian. Note that this choice of eigenstates is analogous to the $|1, 0\rangle \rightarrow |1, -1\rangle$, $D_1$ transition, in the $|F, m_F\rangle$ coupled basis in rubidium-87. The transition strength will be given by

$$|\langle g|er_1|e\rangle|^2$$
$$= a_1^2 b_1^2 |\langle 0, 0, +1/2, -1/2|er_1|1, -1, +1/2, -1/2\rangle|^2. \tag{17}$$

All other terms are zero since the $er_1$ operator can only couple states where the excited $m_L$ quantum number is reduced by one and all other quantum numbers are the same. Since only the orbital angular momentum of the electron is changed during an electric dipole transition, we can decouple the transition into angular and spin parts,

$$|\langle g|er_q|e\rangle|^2 = a_1^2 b_1^2 |\langle L, m_L|er_q|L', m_{L'}\rangle \langle m_S, m_I|m_{S'}, m_{I'}\rangle|^2, \tag{18}$$

where symbols with a prime denote the excited state quantum number. The last part in Eq. (18) ensures that the only non-zero result comes when the electron spin and nuclear spin remain unchanged. Using the Wigner–Eckart theorem [55] we can reduce the angular part,

$$|\langle L, m_L|er_q|L', m_{L'}\rangle|^2 = \begin{pmatrix} L & 1 & L' \\ -m_L & q & m_{L'} \end{pmatrix}^2 \langle L\|e\mathbf{r}\|L'\rangle^2. \tag{19}$$

The symbol in brackets is a Wigner 3-j symbol [56]. When $L' = 1$ and $L = 0$, the square of the Wigner 3-j symbol is 1/3 for any $\sigma^+$, $\sigma^-$ or $\pi$ transition. The double-bar matrix element, $\langle L\|e\mathbf{r}\|L'\rangle$, denotes the reduced dipole matrix element, $d$. Therefore, for our particular example the transition strength is simply,

$$|\langle g|er_1|e\rangle|^2 = \frac{1}{3}d^2 a_1^2 b_1^2 \equiv d^2 c_i^2. \tag{20}$$

Recalling Eq. (7), we see how the strength fits in to the amplitude of the electric susceptibility for a single transition. For the general case we find the strength from $d^2/3$ multiplied by $\sum_{i,j} a_i^2 b_j^2$, where $a_i$ and $b_i$ are the coefficients of two basis states which are allowed by the selection rules.

See Section 3.2 for details of how ElecSus calculates this strength factor using a matrix representation of the Hamiltonian. Note that our program only considers the case where the magnetic field is parallel to the propagation axis of the light, which means that $\pi$ transitions are forbidden [54].

Fig. 1 shows an example for the rubidium $D_1$ line, showing the result of adding the lineshape at each transition frequency, scaled by the relative transition strength.

### 2.3. The Stokes parameters

Together, the four Stokes parameters characterise the polarisation state of light [57]. They are easily measurable with linear optics and photodetectors [58], and are therefore a convenient way to measure the polarisation of light. However, the Stokes parameters have uses beyond measuring polarisation. Predicting their spectral dependence for an atomic medium is useful since they can be used for several optical devices. The $S_0$ parameter, equivalent to transmission, can be used for primary thermometry [23]. The $S_1$ and $S_2$ parameter signals can be used as far off-resonance laser frequency stabilising references [9], while the $S_3$ signal is the dichroic atomic vapour laser lock [59,60] error signal. The Stokes parameters are defined as

$$S_0 \equiv (I_- + I_+) = (I_x + I_y) = (I_\nearrow + I_\searrow), \tag{21a}$$

$$S_1 \equiv (I_x - I_y), \tag{21b}$$

$$S_2 \equiv (I_\nearrow - I_\searrow), \tag{21c}$$

$$S_3 \equiv (I^- - I^+), \tag{21d}$$

where the $I_x$ and $I_y$ are the intensities of light that are transmitted and reflected, respectively, at a polarising beam-splitter placed after the medium; this defines the $x$- and $y$-axes. The symbols $I_\nearrow$ and $I_\searrow$ denote the intensities of light after the medium which are polarised at an angle that deviates from the $x$-axis by 45° and −45° respectively, while $I_-$ and $I_+$ represent the intensities of light which drive $\sigma^\pm$ transitions. ElecSus considers the case of a medium where only $\sigma^\pm$ transitions are allowed. In this case the medium is circularly birefringent and dichroic. It is therefore convenient to parameterise the initial polarisation of the light in terms of the two circular components,

$$\mathbf{E}_0 = E_- \mathbf{e}_- + E_+ \exp(i\phi_0)\mathbf{e}_+, \tag{22}$$

where $\mathbf{E}_0$ is the electric field before the atomic medium and is written as a Jones vector [61,62] in the circular basis, and $\phi_0$ is the phase shift between the two circular components. It is useful to define a quantity ($p$) representing the proportion of the light before the medium that drives $\sigma^-$ transitions,

$$p \equiv \frac{|E_-|^2}{|E_-|^2 + |E_+|^2}. \tag{23}$$

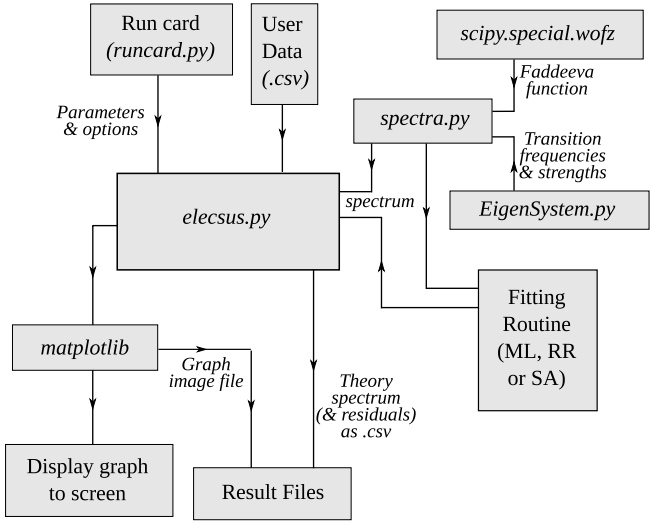

**Fig. 2.** Block diagram showing the flow of information in the ElecSus program.

We can now write the Stokes parameters in terms of $p$ and the extinction coefficients, $\alpha^\pm$, and refractive indices $n^\pm$,

$$S_0 = p \exp(-\alpha_- \ell) + (1-p) \exp(-\alpha_+ \ell), \tag{24a}$$

$$S_1 = 2\sqrt{p-p^2} \cos(2\psi) \exp\left[-\frac{1}{2}(\alpha_- + \alpha_+)\ell\right], \tag{24b}$$

$$S_2 = 2\sqrt{p-p^2} \sin(2\psi) \exp\left[-\frac{1}{2}(\alpha_- + \alpha_+)\ell\right], \tag{24c}$$

$$S_3 = p \exp(-\alpha_- \ell) - (1-p) \exp(-\alpha_+ \ell), \tag{24d}$$

$$\psi = \frac{1}{2}(n_+ - n_-)k\ell + \theta_0, \tag{24e}$$

where $\ell$ is the length of the medium in the propagation direction and $\theta_0 = \phi_0/2$. Note that these are normalised to unit intensity before the medium. For linearly polarised light incident on the medium, $\theta_0$ represents the angle the plane of polarisation makes with the $x$-axis. The parameters $p$ and $\theta_0$ are enough to characterise the initial polarisation state of a coherent laser beam.

We can also calculate the individual $I_x$ and $I_y$ spectra using the definitions (21a) and (21b),

$$I_x = (S_0 + S_1)/2, \tag{25}$$

$$I_y = (S_0 - S_1)/2. \tag{26}$$

These spectra are useful since they correspond to Faraday filtering [16–20].

## 3. Program structure

The program is centred on the `elecsus.py` module. This module takes the user-specified parameters and settings from the run card, imports the `spectrum` function and uses it to generate the spectrum, or imports the fitting modules which fit user data to curves generated by `spectrum`. Experimental data supplied by the user should be as a two-column comma separated values (csv) file; the first column specifies linear detuning in units of GHz while the second is the corresponding value for the spectrum. csv files are readable by spreadsheet programs and are a common data output format on digital storage oscilloscopes. The program then creates a new sub-directory for the output files. It writes the main result to a csv file named by the user-specified label suffixed by `_theory.csv`. If requested in the run card, the module also uses `matplotlib` [63] to plot the curve and display it to the screen; there is also an option to save the plot to a file. If a fit was performed the module will also save the residuals [64] of data and theory curves as a csv file, as well as output the fit parameters to a file suffixed by `_Parameters.txt`. Fig. 2 shows a flowchart of the program structure.

### 3.1. Global lineshape profile

In Section 2.1 we saw that calculating the atomic line-shape, $\mathcal{V}(\Delta_i)$, can be done by performing a convolution. However, the result is related to the Faddeeva function, $w(\zeta)$,

$$\mathcal{V}(\Delta_i) = \frac{\mathrm{i}\sqrt{\pi}}{kU} w(\zeta), \tag{27}$$

$$\zeta = \frac{\mathrm{i}}{2}\frac{\Gamma}{kU} + \frac{\Delta_i}{kU}. \tag{28}$$

Using a `scipy` implementation of the Faddeeva function (`scipy.special.wofz`), which is a wrapper for a fast algorithm written in C++ [65], ElecSus quickly calculates the line-shape. Calculating the global line-shape must be done with care since an inefficient implementation can be slow. The line-shape should be the same for all transitions of the same atom and so the `wofz` function only needs to be called once per isotope. Isotopes of the same element will also have a slightly different line-shape because the different masses provide a different Doppler broadening. The appropriate line-shape profile is then added to the spectrum for each transition with a shift given by the transition detuning and an amplitude proportional to its line-strength.

### 3.2. Matrix representation of the atomic Hamiltonian

For each atomic term ($n^2S_{1/2}$ and $n^2P$) a separate Hamiltonian is built up in a similar manner as described in Section 2.2 as a matrix in the completely uncoupled basis. Each matrix is of size $D_{LSI} \times D_{LSI}$ where $D_{LSI} = (2L+1)(2S+1)(2I+1)$. We set the coarse atomic energy to zero since we are simply looking for the detuning values from a global linecentre. For the $n^2S_{1/2}$ term $L=0$, so there is no fine structure interaction. The Hamiltonian is therefore given by

$$\hat{H}_S = \frac{A_{\mathrm{hf}}}{\hbar^2}(\boldsymbol{I}\cdot\boldsymbol{J}) + \frac{\mu_B B_z}{\hbar}\left(g_S \hat{S}_z + g_I' \hat{I}_z\right). \tag{29}$$

ElecSus calculates D$_1$ and D$_2$ spectra separately, however the full $n^2P$ term is calculated. The eigenenergies of the fine structure part of the Hamiltonian are $-\gamma_\mathrm{f}$ and $\gamma_\mathrm{f}/2$, but since we want to calculate detuning values from a global linecentre we need to re-centre this energy. When addressing the $n^2P_{1/2}$ term we therefore calculate the Hamiltonian as

$$\hat{H}_P = \frac{\gamma_\mathrm{f}}{\hbar^2}(\boldsymbol{L}\cdot\boldsymbol{S}) + \hat{H}_\mathrm{hf} + \hat{H}_Z + \gamma_\mathrm{f}\otimes\mathbb{I}_{D_{LSI}}, \tag{30}$$

where $\mathbb{I}_{D_{LSI}}$ is the identity matrix. If we instead wanted to address the $n^2P_{3/2}$ term we would introduce a $-\gamma_\mathrm{f}/2\otimes\mathbb{I}_{D_{LSI}}$ term. Fig. 3 shows a schematic of the $\hat{H}_P$ matrix after it is diagonalised. When calculating transition strengths, certain parts of the $\hat{H}_P$ matrix are selected depending on whether we want to calculate $\sigma^+$ or $\sigma^-$ transitions, and which D-line. Each strength factor ($C_i$) is then calculated by performing the dot product between one of the rows of $\hat{H}_S$ with one of the post-selected rows of the excited state matrix.

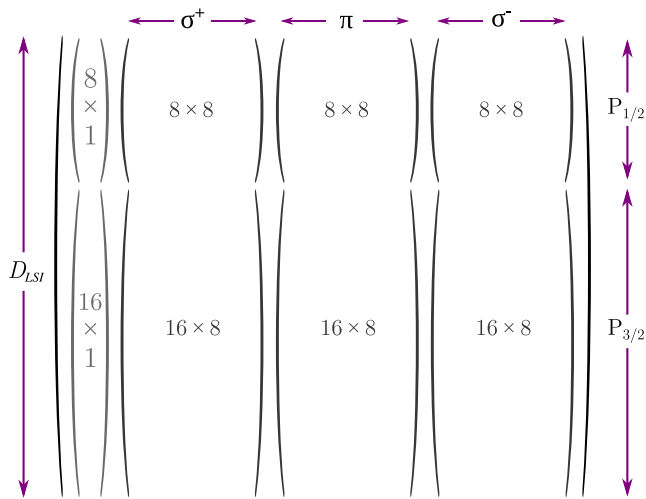

**Fig. 3.** A schematic of the n²P term matrix for ⁸⁷Rb obtained once diagonalised. The first column gives the eigenenergies which are added to the left of the matrix. The corresponding rows give the coefficients of the (completely uncoupled) basis states that make up the eigenstate. Certain parts of the matrix correspond to allowed final states for transitions, and so are post selected when finding transition strengths. In this case the sub-matrices are 8 by 8 for a $D_1$ transition and 16 by 8 for $D_2$.

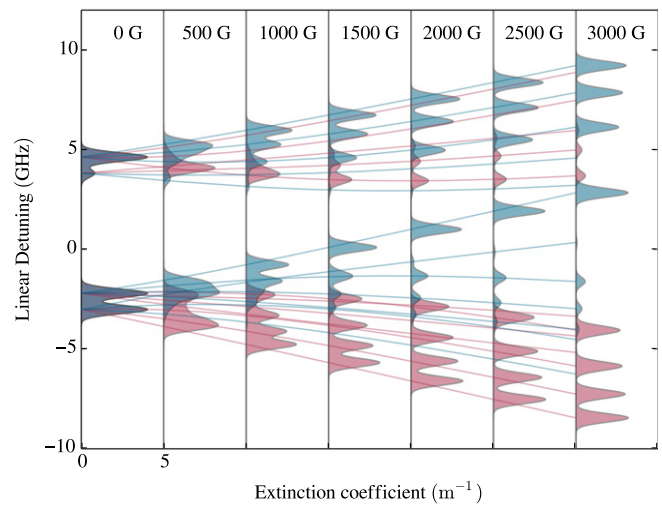

**Fig. 4.** (Colour online) The extinction coefficient as a function of linear detuning ($\Delta/2\pi$) for a ⁸⁷Rb vapour on the $D_1$ line (795 nm) and at a temperature of 20 °C. There are seven plots of the extinction coefficient at magnetic field values ranging from 0 to 3000 G in increments of 500 G. Each plot is placed on its side, displaced to the right for clarity, with the extinction coefficient ranging from 0 to 5 m$^{-1}$ in each one. The area under each curve has been coloured either translucent red or blue, representing $\sigma^-$ and $\sigma^+$ transitions. Overlaying the plots are curves showing the transition frequencies as a function of magnetic field. Notice that the strong $\sigma^-$ transitions move to lower energies while the strong $\sigma^+$ ones move to higher energies. A video showing the evolution of the extinction coefficients for a Caesium vapour is available from http://www.jqc.org.uk/research/project/elecsus/12683.

Many strength factors will be zero (with some numerical computing error) and as such a lower bound on the strength is placed in order to reject these transitions and save computing time. Also, the transition strengths corresponding to $\pi$ transitions are not calculated since they are forbidden. It should be noted that in the case of a zero magnetic field ElecSus adds a very small magnetic field ($10^{-4}$ G) in order to discriminate between the energy levels. This is necessary in order to perform the post-selection. It should also be highlighted that even though we calculate the full P-term Hamiltonian, we cannot use this to calculate both $D_1$ and $D_2$ spectra at the same time, as might be expected. This is because the hyperfine coefficients used for the $n^2P_{1/2}$ term are different to those for the $n^2P_{3/2}$ term. The omission of lithium spectra from the current version of ElecSus is partly due to this reason. ElecSus treats $D_1$ and $D_2$ spectra separately, but lithium $D_1$ and $D_2$ lines are sufficiently close (∼10 GHz) that they need to both be included when predicting a single spectrum.

To demonstrate the power of the Hamiltonian technique, the extinction coefficient calculated by ElecSus is plotted in Fig. 4 for increasing values of the magnetic field. At small and large magnetic field values the spectra are simple, and can be defined by good quantum numbers [39]. In these regimes, it may be sufficient to estimate frequency shifts by a perturbative treatment. However, in the intermediate regime there are no good quantum numbers, and neither the hyperfine interaction nor the magnetic interaction can be considered small. This shows the power of the Hamiltonian technique in being able to find transition frequencies and strengths accurately in all regimes.

### 3.3. Fitting experimental spectra and timing information

The user provides experimental data, in csv format, as two columns of values. The first column specifies the linear detuning (in GHz) while the second gives the spectrum data. If a different linecentre value (from that specified in Appendix B) has been used to make the linear detuning axis, the user can specify a global shift in the run card in order to take this into account.

Fitting theory to experiment involves defining a 'cost' function, which quantifies how far the theory curve deviates from the experimental. This is often defined as the square of the difference between theory and experiment summed at each point along the curve. The cost function is then minimised by changing the parameters which define the theory curve. It is useful to think of this as finding the global minimum in a parameter space. There are three different fitting routines that can be used and should be selected based on the complexity of the fitting problem. A more complex fitting problem will tend to be one with many fit parameters.

In simple cases the option of fitting via the Marquardt–Levenberg (ML) method [64] should be chosen. This method is a 'hill-climbing' algorithm, which will quickly find the local minimum (or maximum, hence the name). Fig. 5 shows four different experimental transmission spectra recorded using a similar technique and apparatus as described in Ref. [15]. The data were taken at different temperatures, and these temperatures were then found by fitting using the ML method. Typically, when fitting one parameter we find no further improvement by using the global fitting routines. One exception to this is the case of fitting high magnetic fields, where the ML technique can fail unless the initial guess is accurate. This is because there may be little or no overlap between the experimental and initial theoretical predictions, creating a plateau in parameter space which is known to be hard for hill-climbing algorithms to deal with [66].

For more complicated fits of more parameters the ML technique may fail to find the global minimum, and so a global fitting routine should be used. Fig. 6 shows the result of fitting three experimental parameters using the ML method and the random-restart [66] (RR) method. The random-restart method is a meta-algorithm which simply performs an ML fit for a range of randomly generated initial states, then picks the best fit. In this way the RR technique has the possibility of escaping the nearest minimum. Another advantage of this method is that it is easily parallelised and can therefore be more time efficient; ElecSus will use all available cores of the computer's CPU in order to perform an RR fit.

Another global fitting routine supplied ElecSus is based on simulated annealing (SA) [67]. It uses the Metropolis algorithm [68] with a $T_{n+1} = T_n/(1 + \delta T_n)$ cooling schedule [69] where $\delta$ is simply a small number. Fig. 7 shows an experimental $S_1$ spectrum fitted to theory with four fit parameters, using both the ML and SA techniques. The ML method clearly does not perform as well as the SA method. ElecSus' RR routine was found to be unreliable when applied to this particular data, but on occasion did manage to find a good fit.

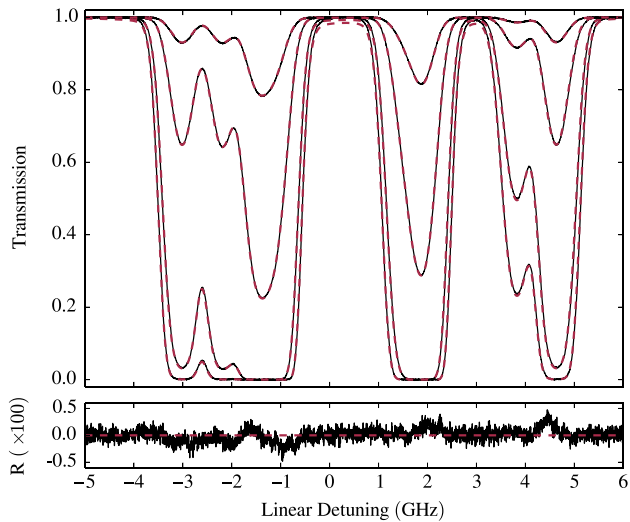

**Fig. 5.** Plots of transmission versus detuning on the $D_1$ line (795 nm) for four different temperatures of a 75 mm long rubidium vapour cell with natural isotopic abundance. The (black) solid line shows experimental data while the (red) dashed curves are theoretical results using ElecSus with temperature as the only fit parameter. Lower curves correspond to higher temperatures; the temperatures were found to be 18.6 °C, 36.1 °C, 59.8 °C and 69.3 °C. Underneath, the residuals (R) between experiment and theory are plotted for the 18.6 °C data set (rms deviation of 0.1%). The magnetic field was assumed to be zero.

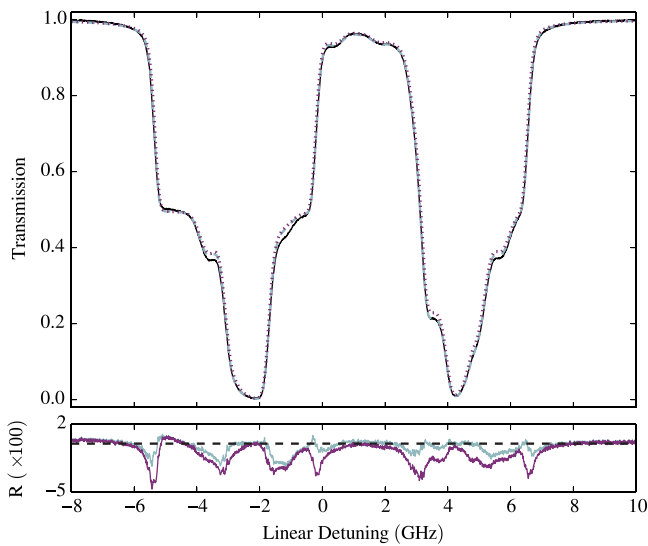

**Fig. 6.** (Colour online) Transmission as a function of linear detuning on the $D_1$ line (795 nm) for a $^{87}$Rb cell (99.0% $^{87}$Rb, 1.0% $^{85}$Rb) with a length of 1 mm. The solid (black) curve shows the experimental data while the dashed (blue) and dotted (purple) curves show fits to the theory using the RR and ML fitting routines. rms deviations from the experimental data of 0.5% and 1% were found for the RR and ML fits respectively. The fit parameters were the magnetic field, cell temperature, and increased Lorentzian broadening due to buffer gases.

To compare the speed of the different fitting routines the data provided in Fig. 6 was fit to theory using all three fitting routines, with the same precision (10 MHz) and the same number of points (4774). A computer with an Intel®Core™ i3-3220 processor was used. The time taken was found to be 13 s, 80 s and 10 min for the ML, RR and SA techniques respectively. Note that the RR technique involves evaluating spectra in parallel (4 parallel processes in this case) whilst the other two are sequential.

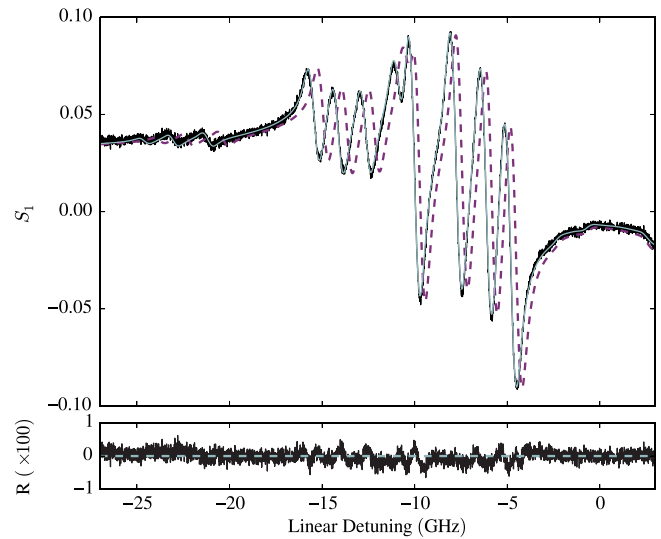

**Fig. 7.** (Colour online) $S_1$ Stokes parameter as a function of linear detuning on the $D_2$ line (780 nm) for a $^{87}$Rb cell (99.0% $^{87}$Rb, 1.0% $^{85}$Rb) with a length of 1 mm. The solid (black) line is the experimental data while the solid (blue) and dashed (purple) lines are the fits to theory using the SA and ML techniques respectively. The ML fit is clearly poor; shown underneath are the residuals for the SA fit. The rms deviations of the ML and SA fits were 2.2% and 0.16% respectively.

### 3.3.1. Uncertainties in fit parameters

When fitting experiment to theory a $\chi^2$ analysis [64] is often used to extract the statistical uncertainty in the fit parameters. This requires knowledge of the uncertainty of the data points which will vary from experiment to experiment. The current version of ElecSus has no facility to accept these uncertainties as an input and so does not provide uncertainties in the fit parameters. However, since ElecSus outputs the theoretical curve as csv file, the user can perform this analysis manually (see Ref. [64]).

Another method to extract the statistical uncertainty in the fit parameters is to take several data sets under nominally the same experimental conditions and fit them independently. Then the mean and standard error of each parameter can be found. By this method we have found that our experimental technique typically gives statistical uncertainties of ∼0.1 °C for temperature and ∼1 G for magnetic field [10] (at large magnetic field values).

### 4. Installation and usage

The program can be downloaded from http://www.jqc.org.uk/research/project/elecsus/12683 as a zip file named `ElecSus.zip`. Installation should be complete after simply extracting the `ElecSus.zip` file. No further action is required since it is assumed that the prerequisites (NumPy, SciPy and Matplotlib) are installed, and are found by the python interpreter. To use the program, the following steps should be taken.

1. Open a command prompt window and move to the directory extracted from the zip file (main directory).
2. Copy the file `runcard.py` to create your own file `myruncard.py`.
3. If fitting is required, place the csv file containing the data in the main directory.
4. Change the values and options in the file `myruncard.py` to the desired ones.
5. Execute the following command in the prompt window:
   `python elecsus.py myruncard.py`
   If the `myruncard.py` option is omitted the program will, by default, look in `runcard.py` for parameter values and options.

### 4.1. Run card parameters and options

In the run card the user must specify which alkali element, D-line and spectrum they want to compute. The choices of spectra for ElecSus to compute are the refractive indices, $n^{\pm}$, the group indices, $n_g^{\pm}$, Stokes parameters and the individual $I_x$ and $I_y$ spectra. The user can then specify parameters such as temperature, cell length etc., and whether they would like to fit experimental data or simply produce a theoretical prediction. The variables in the run card are python variables, and as such the order in which they are specified does not matter.

We should note that there are two temperature variables, one which parameterises the number density (see Appendix A) and one which parameterises the Doppler broadening, where the second is constrained to the first by default. The option to treat these two temperatures separately is included for two reasons. The first reason is that non-uniform heating of vapour cells can cause the vapour temperature near the laser beam to be different to the temperature at or near the metal reservoir. The second reason is that the number density formulas used are only quoted to be accurate to about 5% [70], and so when fitting to a sufficiently accurate and precise experimental spectrum we would expect the two temperatures to disagree even if the whole cell is in thermal equilibrium. This disagreement would correspond to the systematic error in the vapour pressure formula.

It should also be noted that not all the options are relevant, and in these cases ElecSus will ignore those entries. For example when fitting to data the options that define the detuning axis are ignored since this is defined by the first column of the csv data file provided by the user. However, these variables should not be deleted from the run card.

### 4.2. Test runs

Provided with ElecSus are two sets of experimental data taken using the same experimental procedure as described in Ref. [10]. Also provided are two sample run cards, which can immediately be used by the user to ensure that the program is working on their computer.

The first example can be run by executing the following command in a prompt window whilst in the main directory:

```
python elecsus.py runcard_D1sample.py
```

This will use the data provided in the data file `SampleDataRbD1.csv`, and then fit three parameters using the RR method. Since this data is the same as shown in Fig. 6, and the same parameters are being fitted, the result should be very similar to that found in Fig. 6. Fitting should not take longer than a few minutes.

In a similar way the user can run the second example with the `runcard_D2sample.py` file. This will use data from the `SampleDataRbD2.csv` file, which is the same as the data shown in Fig. 7. Running this example will fit the data using the SA technique and should give a correspondingly similar result to that seen in the figure. Note that this example can take up to 30 min to run.

## 5. Example applications

Here we show two applications of ElecSus; Faraday filtering and pulse propagation.

### 5.1. Faraday filters

Faraday filters are made by placing a Faraday rotator between two crossed polarisers [16]. Using an atomic medium as the Faraday rotator means typically only frequencies close to a resonance line are rotated and hence creating an ultranarrow filter. Here we predict Faraday filtering spectra on the $D_2$ line for all four alkali-metals programmed in ElecSus.

To generate a Faraday filtering spectrum in ElecSus we first need to emulate the effect of the first polarising beam splitter cube by setting $p$ to 50% (see Eq. (22)) and $\theta_0$ to 90°. This defines the light before the medium as linearly polarised in the $y$-direction. We then choose to plot the $I_x$ spectrum which corresponds to light transmitted through a second polariser crossed with the first. Note that choosing $\theta_0 = 0$ and plotting $I_y$ gives the same result.

The spectral profile of the filter can be controlled by changing cell length, magnetic field and temperature. The optimal spectral profile is dependent on the application; we have chosen to emulate filters in the line-centre operation [18]. Fig. 8 shows the results for a 75 mm long atomic medium.

### 5.2. Pulse propagation

To highlight the power of ElecSus we follow an example where we have used it to design a laser frequency stabilising signal and also extract key information about the experimental cell in order to accurately predict weak-probe pulse propagation. To demonstrate this a 2 mm long rubidium vapour cell with natural isotopic abundance was heated, and then nanosecond long pulses (generated using a Pockels cell between two crossed polarisers) were sent through the medium. This experiment is very similar to that performed in Ref. [46].

To find how a pulse propagates through the medium we must know how the transmission and refractive index of the medium vary across the bandwidth of the pulse. A Fourier transform of the temporal pulse profile gives the pulse profile in terms of frequency, and then the effect of transmission and phase shifts can be applied. An inverse Fourier transform then yields the pulse after traversing the medium.

To measure the required transmission and refractive index spectra, a weak continuous laser beam was sent through the medium and the transmission of the beam was measured as a function of laser frequency (see inset (a) of Fig. 9). ElecSus was then used to fit this spectrum to find the two temperatures that parameterise number density and the Doppler broadening. These were found to be 179.4 °C and 189.6 °C respectively, which may indicate asymmetric heating of the cell. ElecSus can then be used again to infer what the refractive index of the medium should be given these parameters.

To measure the pulse profile accurately enough, single photon counting modules with good timing resolution were used. To build up an acceptable profile, counting was done over ∼15 min for approximately 100k repetitions. Over this length of time the laser frequency needs to be actively stabilised in order to prevent the carrier frequency of the pulse from drifting. We use a technique similar to that described in Ref. [9]. The experimental parameters required to give suitable frequency references were found using ElecSus. We found that using the $S_1$ Stokes parameter with temperatures around ∼140 °C and magnetic fields around 200 G, gave zero crossings both far off resonance and near zero detuning when using a 75 mm long rubidium vapour cell. This is useful since the near zero detuning reference can be used to demonstrate slow-light while the far off resonance reference is a good approximation to the reference pulse (defined as a pulse that traversed vacuum instead of the atomic medium). Inset (b) of Fig. 9 shows the raw photodetector signal used as the frequency references.

Fig. 9 shows the experimentally measured pulse after the medium and the theoretical prediction. We can see that the theoretical prediction matches the experimental data within the precision of the experiment; no fit parameters are used. The pulse shows a large time delay which is accurately predicted by the theory.

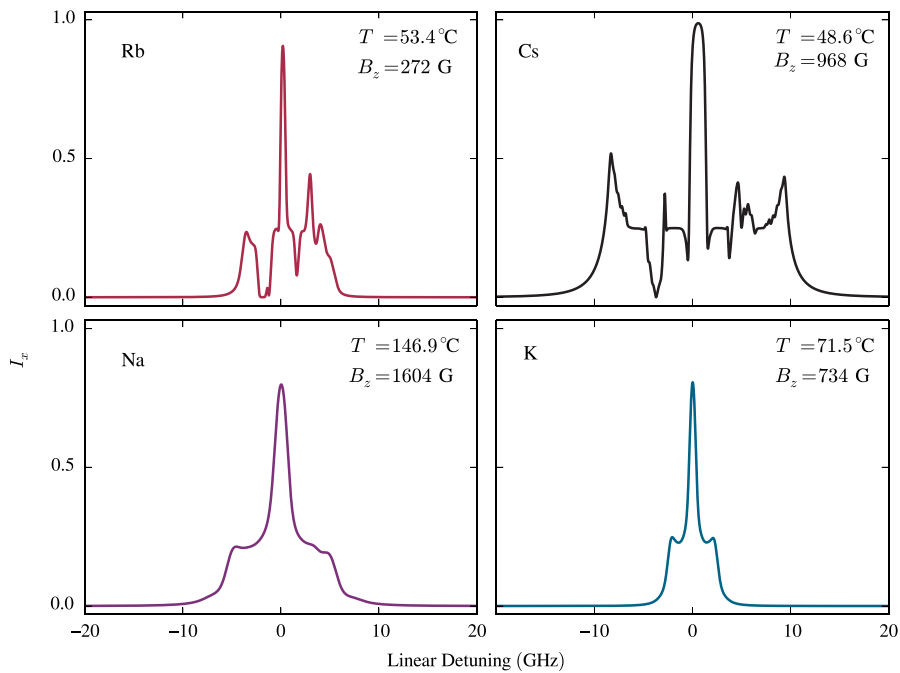

**Fig. 8.** Faraday filtering. The intensity of light polarised in the $x$-direction ($I_x$) after propagating through a 75 mm long atomic medium. The incident intensity is normalised to one and the light is linearly polarised ($p = 50\%$) in the $y$-direction ($\theta_0 = 90°$). The vertical axis of each plot ranges from 0 to 1. Each horizontal axis shows the linear detuning from the $D_2$ global linecentre of each element (see Appendix B), ranging from $-20$ to $20$ GHz. The temperatures parameterising number density and Doppler broadening were constrained to be the same, while isotopic ratios were set to their natural values.

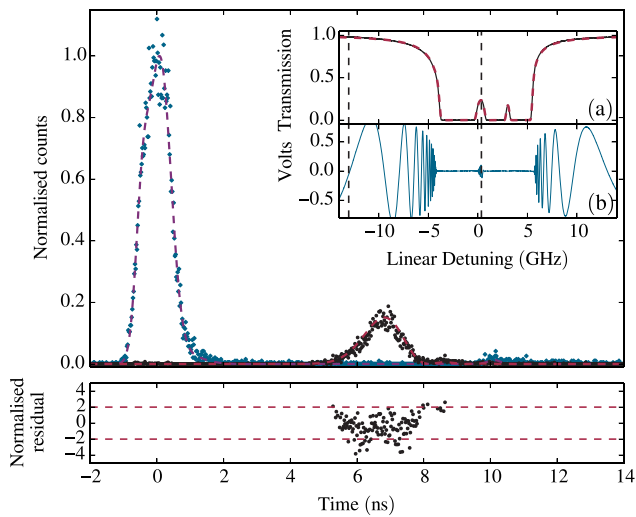

**Fig. 9.** (Colour online) Slowlight pulse propagation through a 2 mm long rubidium vapour at natural abundance on the $D_1$ line (795 nm). The (black) dots and (blue) diamonds show the number of photons counted as a function of time (20 ps bin width) for a laser pulse with a carrier detuning near and far from resonance respectively. The counting was completed after approximately 100 k repetitions. The (purple) dashed line shows a multiple-Gaussian fit used to characterise the reference pulse. The (red) dashed line shows the theoretical prediction of pulse intensity measured after the cell. The data is normalised to the peak intensity of the reference pulse, while the time axis is centred at this peak. Underneath the graph are plotted the normalised residuals [64], which show good agreement (bins with less than 5 counts were omitted since they have been judged statistically insignificant [64]). Inset (a) shows a theoretical fit to a CW experimental transmission scan, performed immediately after the pulse experiment in order to extract the number density temperature and Doppler temperature of the medium. These are taken to be 179.4 °C and 189.6 °C respectively. Inset (b) shows a Faraday signal from a 75 mm long cell used to stabilise the frequency of the laser [9] during the experiment. The black vertical dashed lines show the locking points used to stabilise the carrier detuning to $(354 \pm 2)$ MHz, and to $(-13.0 \pm 0.1)$ GHz when measuring the reference pulse.

This gives evidence that the refractive index spectrum predicted by ElecSus is accurate.

## 6. Conclusions and outlook

We have presented a computer program to calculate the electric susceptibility of an alkali-metal vapour, and we describe the underlying model used. The program can be used to design optical devices such as Faraday filters, and laser frequency stabilising references. The fitting routines provided in the program allow the user to measure experimental parameters (such as temperature or magnetic field).

We have modelled the D-lines of Na, K, Rb and Cs; a future version will include Li. Also, extending the program beyond D-line transitions will prove useful. For example, including transitions from the $n$S ground state to the $(n + 1)$P excited states will be useful for modelling Faraday filtering on these lines [71] as well as experiments utilising these transitions for creating high phase-space density magneto-optical traps [72–74]. Also, extending the model to include transitions between excited states may allow modelling of excited state Faraday filters [75–77].

Working with atoms in confined geometries such as nanometric thin cells [78] or hollow-core fibres [79–84], introduces deviations of the atomic lineshape from the simple Voigt profile. This is due to effects such as atom–surface interactions [85,86,40] and Dicke narrowing. Extending the model to allow different atomic lineshapes has already been shown to accurately account for these effects [1,40], and so we intend to extend ElecSus to include these effects.

Fig. 9 motivates future versions of the program to contain a pulse propagation feature. It again shows that the medium can be accurately probed at GHz bandwidth [87]. As such fitting to short pulses should allow further improvements to accuracy and facilitate observation of dynamics on short time-scales [88].

### Acknowledgements

We thank K.A. Whittaker, R.S. Mathew, W.J. Hamlyn and N. Sibalic for testing the code and providing feedback. We would also like to thank R.M. Potvliege for reading and commenting on the manuscript. We acknowledge financial support from EPSRC (grant EP/L023024/1) and Durham University. The data presented in this paper are available upon request.

**Table A.1**
Values of the constants used in Eq. (A.1) to determine the vapour pressure. Values are taken from [70].

| Element (phase) | A | B | C |
|---|---|---|---|
| Na (solid) | 5.298 | −5603 | |
| Na (liquid) | 8.400 | −5634 | −1.1748 |
| K (solid) | 4.961 | −4646 | |
| K (liquid) | 8.233 | −4693 | −1.2403 |
| Rb (solid) | 4.857 | −4215 | |
| Rb (liquid) | 8.316 | −4275 | −1.3102 |
| Cs (solid) | 4.711 | −3999 | |
| Cs (liquid) | 8.232 | −4062 | −1.3359 |

**Table A.2**
Natural abundances of sodium, potassium, rubidium and caesium. Taken from [89].

| Element | Mass number | Abundance (%) |
|---|---|---|
| Na | 23 | 100 |
| K | 39 | 93.2581 |
| | 40 | 0.0117 |
| | 41 | 6.7302 |
| Rb | 85 | 72.17 |
| | 87 | 27.83 |
| Cs | 133 | 100 |

## Appendix A. Vapour pressure equations and isotopic abundances

Here we list the vapour pressure equations that are used to find the atomic number density as a function of temperature, $T$. The equations for all four elements are of the form

$$\log_{10}(p[\text{atm}]) = A + B(T[\text{K}])^{-1} + C \log_{10}(T[\text{K}]), \quad (A.1)$$

where $p$ is the vapour pressure, values in square brackets denote the units of the corresponding variables and $A$, $B$ and $C$ are constants specific to the element and its phase (solid or liquid). Table A.1 shows the values of the constants used for the ElecSus program. All the constants were taken from the "precise" values given in [70].

The atomic number density can then be found from the pressure by assuming an ideal gas,

$$\mathcal{N} = \frac{p}{k_B T}. \quad (A.2)$$

The conversion factor for atmospheres to pascals (1 atm = 101 325 Pa) is also used in order to get the number density in units of atoms/m$^3$. It should be noted that these equations will give the elemental atomic number density and so they need to be reduced by the appropriate factor when considering one isotope of the element. For example the number density of $^{87}$Rb in a naturally abundant vapour will be given by the Rb elemental number density multiplied by 0.2783. Table A.2 shows the natural isotopic abundances.

## Appendix B. Physical constants

This section lists the values of the physical constants used by ElecSus (see Tables B.3–B.12). Note that there has been no attempt to reduce floating point rounding error in calculations. References denote either where the constants have been found explicitly or for where numbers were found in order to calculate the given constant. The user can change these numbers by changing the entries in the `FundamentalConstants.py` and `AtomConstants.py` files. Isotope shifts are given relative to the linecentre values; negative isotope shifts denote an increase in the hyperfine free transition frequency. Uncertainties in the values are only given for completeness and do not feature in the code.

**Table B.3**
Fundamental constants which are loaded from SciPy v0.13.2 library [90], which in turn come from the 2010 CODATA recommended values [91]. Updating SciPy may automatically update these physical constants for ElecSus.

| Quantity | Symbol | Value |
|---|---|---|
| Electron spin $g$-factor | $g_S$ | 2.00231930436153(53) |
| Bohr magneton | $\mu_B$ | $9.27400968(20) \times 10^{-24}$ JT$^{-1}$ |
| Boltzmann constant | $k_B$ | $1.3806488(13) \times 10^{-23}$ JK$^{-1}$ |
| Atomic mass unit | $u$ | $1.660538921(73) \times 10^{-27}$ kg |
| Vacuum permittivity | $\epsilon_0$ | $8.854187817620389 \times 10^{-12}$ Fm$^{-1}$ |
| Bohr radius | $a_0$ | $5.2917721092(17) \times 10^{-11}$ m |
| Reduced Planck constant | $\hbar$ | $1.054571726(47) \times 10^{-34}$ Js |
| Speed of light | $c$ | 299792458 ms$^{-1}$ |
| Elementary charge | $e$ | $1.602176565(35) \times 10^{-19}$ C |

**Table B.4**
Constants related to the D$_1$ line. The linecentre refers to the definition of zero global detuning.

| Element | Quantity | Symbol | Value | Reference |
|---|---|---|---|---|
| Na | Linecentre wavelength | $\lambda_0$ | 589.7558147(15) nm | [92] |
| | Linecentre frequency | $\nu_0$ | 508.3331958(13) THz | [92] |
| | Natural linewidth (FWHM) | $\Gamma_0$ | $2\pi \times 9.765(13)$ MHz | [93] |
| K | Linecentre wavelength | $\lambda_0$ | 770.10836827(12) nm | [94] |
| | Linecentre frequency | $\nu_0$ | 389.286067199(63) THz | [94] |
| | Natural linewidth (FWHM) | $\Gamma_0$ | $2\pi \times 5.956(11)$ MHz | [95] |
| Rb | Linecentre wavelength | $\lambda_0$ | 794.978969380(82) nm | [96] |
| | Linecentre frequency | $\nu_0$ | 377.107407299(39) THz | [96] |
| | Natural linewidth (FWHM) | $\Gamma_0$ | $2\pi \times 5.746(8)$ MHz | [97] |
| Cs | Linecentre wavelength | $\lambda_0$ | 894.59325986(11) nm | [98] |
| | Linecentre frequency | $\nu_0$ | 335.116048807(41) THz | [98] |
| | Natural linewidth (FWHM) | $\Gamma_0$ | $2\pi \times 4.584(8)$ MHz | [99] |

**Table B.5**
Constants related to the $D_2$ line. The linecentre refers to the definition of zero global detuning.

| Element | Quantity | Symbol | Value | Reference |
|---|---|---|---|---|
| Na | Linecentre wavelength | $\lambda_0$ | 589.1583264(15) nm | [92] |
|  | Linecentre frequency | $\nu_0$ | 508.8487162(13) THz | [92] |
|  | Natural linewidth (FWHM) | $\Gamma_0$ | $2\pi \times 9.792(13)$ MHz | [93] |
| K | Linecentre wavelength | $\lambda_0$ | 766.70090511(24) nm | [94] |
|  | Linecentre frequency | $\nu_0$ | 391.01617854(12) THz | [94] |
|  | Natural linewidth (FWHM) | $\Gamma_0$ | $2\pi \times 6.035(11)$ MHz | [95] |
| Rb | Linecentre wavelength | $\lambda_0$ | 780.24132411(2) nm | [100,101] |
|  | Linecentre frequency | $\nu_0$ | 384.23042812(1) THz | [100,101] |
|  | Natural linewidth (FWHM) | $\Gamma_0$ | $2\pi \times 6.065(9)$ MHz | [97] |
| Cs | Linecentre wavelength | $\lambda_0$ | 852.34727582(27) nm | [102] |
|  | Linecentre frequency | $\nu_0$ | 351.72571850(11) THz | [102] |
|  | Natural linewidth (FWHM) | $\Gamma_0$ | $2\pi \times 5.225(7)$ MHz | [103] |

**Table B.6**
Constants related to the $^{23}$Na atom.

| Quantity | Symbol | Value | Reference |
|---|---|---|---|
| Nuclear spin | $I$ | 3/2 | |
| Nuclear spin $g$-factor | $g_I'$ | −0.0008046108(8) | [104] |
| Magnetic dipole constant for $5^2S_{1/2}$ | $A_{hf}$ | 885.8130644(5) MHz·$h$ | [104] |
| Magnetic dipole constant for $5^2P_{1/2}$ | $A_{hf}$ | 94.44(13) MHz·$h$ | [105] |
| Magnetic dipole constant for $5^2P_{3/2}$ | $A_{hf}$ | 18.534(15) MHz·$h$ | [106] |
| Electric quadrupole constant for $5^2P_{3/2}$ | $B_{hf}$ | 2.724(30) MHz·$h$ | [106] |
| Mass | $m$ | 22.9897692807(28) $u$ | [107] |

**Table B.7**
Constants related to the $^{39}$K atom.

| Quantity | Symbol | Value | Reference |
|---|---|---|---|
| Nuclear spin | $I$ | 3/2 | |
| Nuclear spin $g$-factor | $g_I'$ | −0.00014193489(12) | [104] |
| Magnetic dipole constant for $5^2S_{1/2}$ | $A_{hf}$ | 230.8598601(3) MHz·$h$ | [104] |
| Magnetic dipole constant for $5^2P_{1/2}$ | $A_{hf}$ | 27.775(42) MHz·$h$ | [94] |
| Magnetic dipole constant for $5^2P_{3/2}$ | $A_{hf}$ | 6.093(25) MHz·$h$ | [94] |
| Electric quadrupole constant for $5^2P_{3/2}$ | $B_{hf}$ | 2.786(71) MHz·$h$ | [94] |
| Isotope shift ($D_1$) | $E_{iso}$ | 8.483 MHz·$h$ | [94] |
| Isotope shift ($D_2$) | $E_{iso}$ | 8.51 MHz·$h$ | [94] |
| Mass | $m$ | 38.96370668(20) $u$ | [108] |

**Table B.8**
Constants related to the $^{40}$K atom.

| Quantity | Symbol | Value | Reference |
|---|---|---|---|
| Nuclear spin | $I$ | 4 | |
| Nuclear spin $g$-factor | $g_I'$ | 0.000176490(34) | [104] |
| Magnetic dipole constant for $5^2S_{1/2}$ | $A_{hf}$ | −285.7308(24) MHz·$h$ | [104] |
| Magnetic dipole constant for $5^2P_{1/2}$ | $A_{hf}$ | −34.523(25) MHz·$h$ | [94] |
| Magnetic dipole constant for $5^2P_{3/2}$ | $A_{hf}$ | −7.585(10) MHz·$h$ | [94] |
| Electric quadrupole constant for $5^2P_{3/2}$ | $B_{hf}$ | −3.445(90) MHz·$h$ | [94] |
| Isotope shift ($D_1$) | $E_{iso}$ | −117.154 MHz·$h$ | [94] |
| Isotope shift ($D_2$) | $E_{iso}$ | −117.51 MHz·$h$ | [94] |
| Mass | $m$ | 39.96399848(21) $u$ | [108] |

**Table B.9**
Constants related to the $^{41}$K atom.

| Quantity | Symbol | Value | Reference |
|---|---|---|---|
| Nuclear spin | $I$ | 3/2 | |
| Nuclear spin $g$-factor | $g_I'$ | −0.00007790600(8) | [104] |
| Magnetic dipole constant for $5^2S_{1/2}$ | $A_{hf}$ | 127.0069352(6) MHz·$h$ | [104] |
| Magnetic dipole constant for $5^2P_{1/2}$ | $A_{hf}$ | 15.245(42) MHz·$h$ | [94] |
| Magnetic dipole constant for $5^2P_{3/2}$ | $A_{hf}$ | 3.363(25) MHz·$h$ | [94] |
| Electric quadrupole constant for $5^2P_{3/2}$ | $B_{hf}$ | 3.351(71) MHz·$h$ | [94] |
| Isotope shift ($D_1$) | $E_{iso}$ | −227.006 MHz·$h$ | [94] |
| Isotope shift ($D_2$) | $E_{iso}$ | −227.67 MHz·$h$ | [94] |
| Mass | $m$ | 40.96182576(21) $u$ | [108] |

**Table B.10**
Constants related to the $^{85}$Rb atom.

| Quantity | Symbol | Value | Reference |
|---|---|---|---|
| Nuclear spin | $I$ | 5/2 | |
| Nuclear spin $g$-factor | $g_I'$ | −0.0002936400(6) | [104] |
| Magnetic dipole constant for $5^2S_{1/2}$ | $A_{hf}$ | 1011.910813(2) MHz·$h$ | [104] |
| Magnetic dipole constant for $5^2P_{1/2}$ | $A_{hf}$ | 120.640(20) MHz·$h$ | [96] |
| Magnetic dipole constant for $5^2P_{3/2}$ | $A_{hf}$ | 25.038(5) MHz·$h$ | [109] |
| Electric quadrupole constant for $5^2P_{3/2}$ | $B_{hf}$ | 26.011(22) MHz·$h$ | [109] |
| Isotope shift (D$_1$) | $E_{iso}$ | 21.624 MHz·$h$ | [96] |
| Isotope shift (D$_2$) | $E_{iso}$ | 21.734 MHz·$h$ | [101] |
| Mass | $m$ | 84.911789732(14) $u$ | [107] |

**Table B.11**
Constants related to the $^{87}$Rb atom.

| Quantity | Symbol | Value | Reference |
|---|---|---|---|
| Nuclear spin | $I$ | 3/2 | |
| Nuclear spin $g$-factor | $g_I'$ | −0.0009951414(10) | [104] |
| Magnetic dipole constant for $5^2S_{1/2}$ | $A_{hf}$ | 3417.34130545215(5) MHz·$h$ | [110] |
| Magnetic dipole constant for $5^2P_{1/2}$ | $A_{hf}$ | 406.147(15) MHz·$h$ | [96] |
| Magnetic dipole constant for $5^2P_{3/2}$ | $A_{hf}$ | 84.7185(20) MHz·$h$ | [100] |
| Electric quadrupole constant for $5^2P_{3/2}$ | $B_{hf}$ | 12.4965(37) MHz·$h$ | [100] |
| Isotope shift (D$_1$) | $E_{iso}$ | −56.077 MHz·$h$ | [96] |
| Isotope shift (D$_2$) | $E_{iso}$ | −56.361 MHz·$h$ | [101] |
| Mass | $m$ | 86.909180520(15) $u$ | [107] |

**Table B.12**
Constants related to the $^{133}$Cs atom.

| Quantity | Symbol | Value | Reference |
|---|---|---|---|
| Nuclear spin | $I$ | 7/2 | |
| Nuclear spin $g$-factor | $g_I'$ | −0.00039885395(52) | [104] |
| Magnetic dipole constant for $5^2S_{1/2}$ | $A_{hf}$ | 2298.1579425 MHz·$h$ (exact) | [104] |
| Magnetic dipole constant for $5^2P_{1/2}$ | $A_{hf}$ | 291.922(20) MHz·$h$ | [102] |
| Magnetic dipole constant for $5^2P_{3/2}$ | $A_{hf}$ | 50.28827(23) MHz·$h$ | [53] |
| Electric quadrupole constant for $5^2P_{3/2}$ | $B_{hf}$ | −0.4934(17) MHz·$h$ | [53] |
| Mass | $m$ | 132.905451933(24) $u$ | [108] |

## References

[1] J. Keaveney, A. Sargsyan, U. Krohn, I.G. Hughes, D. Sarkisyan, C.S. Adams, Phys. Rev. Lett. 108 (2012) 173601. http://dx.doi.org/10.1103/PhysRevLett.108.173601.
[2] A. Sargsyan, G. Hakhumyan, C. Leroy, Y. Pashayan-Leroy, A. Papoyan, D. Sarkisyan, Opt. Lett. 37 (2012) 1379–1381. http://dx.doi.org/10.1364/OL.37.001379.
[3] B. Julsgaard, A. Kozhekin, E.S. Polzik, Nature 413 (2001) 400–403. http://dx.doi.org/10.1038/35096524.
[4] U. Vogl, M. Weitz, Nature 461 (2009) 70–73. http://dx.doi.org/10.1038/nature08203.
[5] C. Carr, R. Ritter, C.G. Wade, C.S. Adams, K.J. Weatherill, Phys. Rev. Lett. 111 (2013) 113901. http://dx.doi.org/10.1103/PhysRevLett.111.113901.
[6] P.D.D. Schwindt, S. Knappe, V. Shah, L. Hollberg, J. Kitching, L.-A. Liew, J. Moreland, Appl. Phys. Lett. 85 (2004) 6409. http://dx.doi.org/10.1063/1.1839274.
[7] S. Knappe, V. Shah, P.D.D. Schwindt, L. Hollberg, J. Kitching, L.-A. Liew, J. Moreland, Appl. Phys. Lett. 85 (2004) 1460. http://dx.doi.org/10.1063/1.1787942.
[8] C. Affolderbach, G. Mileti, Rev. Sci. Instrum. 76 (2005) 073108. http://dx.doi.org/10.1063/1.1979493.
[9] A.L. Marchant, S. Händel, T.P. Wiles, S.A. Hopkins, C.S. Adams, S.L. Cornish, Opt. Lett. 36 (2011) 64–66. http://dx.doi.org/10.1364/OL.36.000064.
[10] M.A. Zentile, R. Andrews, L. Weller, S. Knappe, C.S. Adams, I.G. Hughes, J. Phys. B: At. Mol. Opt. Phys. 47 (2014) 075005. http://dx.doi.org/10.1088/0953-4075/47/7/075005.
[11] A. Vernier, S. Franke-Arnold, E. Riis, A.S. Arnold, Opt. Express 18 (2010) 17020–17026. http://dx.doi.org/10.1364/OE.18.017020.
[12] G. Walker, A.S. Arnold, S. Franke-Arnold, Phys. Rev. Lett. 108 (2012) 243601. http://dx.doi.org/10.1103/PhysRevLett.108.243601.
[13] B. Julsgaard, J. Sherson, J.I. Cirac, J. Fiurášek, E.S. Polzik, Nature 432 (2004) 482–486. http://dx.doi.org/10.1038/nature03064.
[14] D.G. England, P.S. Michelberger, T.F.M. Champion, K.F. Reim, K.C. Lee, M.R. Sprague, X.-M. Jin, N.K. Langford, W.S. Kolthammer, J. Nunn, I.A. Walmsley, J. Phys. B: At. Mol. Opt. Phys. 45 (2012) 124008. http://dx.doi.org/10.1088/0953-4075/45/12/124008.
[15] P. Siddons, C.S. Adams, C. Ge, I.G. Hughes, J. Phys. B: At. Mol. Opt. Phys. 41 (2008) 155004. http://dx.doi.org/10.1088/0953-4075/41/15/155004.
[16] D.J. Dick, T.M. Shay, Opt. Lett. 16 (1991) 867–869. http://dx.doi.org/10.1364/OL.16.000867.
[17] J. Menders, K. Benson, S.H. Bloom, C.S. Liu, E. Korevaar, Opt. Lett. 16 (1991) 846–848. http://dx.doi.org/10.1364/OL.16.000846.
[18] H. Chen, C.Y. She, P. Searcy, E. Korevaar, Opt. Lett. 18 (1993) 1019–1021. http://dx.doi.org/10.1364/OL.18.001019.
[19] R.P. Abel, U. Krohn, P. Siddons, I.G. Hughes, C.S. Adams, Opt. Lett. 34 (2009) 3071–3073. http://dx.doi.org/10.1364/OL.34.003071.
[20] J.A. Zielińska, F.A. Beduini, N. Godbout, M.W. Mitchell, Opt. Lett. 37 (2012) 524–526. http://dx.doi.org/10.1364/OL.37.000524.
[21] L. Weller, K.S. Kleinbach, M.A. Zentile, S. Knappe, I.G. Hughes, C.S. Adams, Opt. Lett. 37 (2012) 3405–3407. http://dx.doi.org/10.1364/OL.37.003405.
[22] D. Budker, M. Romalis, Nat. Phys. 3 (2007) 227–234. http://dx.doi.org/10.1038/nphys566.
[23] G. Casa, A. Castrillo, G. Galzerano, R. Wehr, A. Merlone, D. Di Serafino, P. Laporta, L. Gianfrani, Phys. Rev. Lett. 100 (2008) 200801. http://dx.doi.org/10.1103/PhysRevLett.100.200801.
[24] M. Triki, C. Lemarchand, B. Darquié, P.L.T. Sow, V. Roncin, C. Chardonnet, C. Daussy, Phys. Rev. A 85 (2012) 062510. http://dx.doi.org/10.1103/PhysRevA.85.062510.
[25] G.-W. Truong, J.D. Anstie, E.F. May, T.M. Stace, A.N. Luiten, Phys. Rev. A 86 (2012) 030501. http://dx.doi.org/10.1103/PhysRevA.86.030501.
[26] Z. Wu, M. Kitano, W. Happer, M. Hou, J. Daniels, Appl. Opt. 25 (1986) 4483–4492. http://dx.doi.org/10.1364/AO.25.004483.
[27] E. Vliegen, S. Kadlecek, L.W. Anderson, T.G. Walker, C.J. Erickson, W. Happer, Nucl. Instrum. Methods A 460 (2001) 444–450. http://dx.doi.org/10.1016/S0168-9002(00)01061-5.
[28] U.D. Rapol, A. Wasan, V. Natarajan, Phys. Rev. A 64 (2001) 023402. http://dx.doi.org/10.1103/PhysRevA.64.023402.
[29] K.L. Moore, T.P. Purdy, K.W. Murch, S. Leslie, S. Gupta, D.M. Stamper-Kurn, Rev. Sci. Instrum. 76 (2005) 023106. http://dx.doi.org/10.1063/1.1841852.
[30] D.R. Scherer, D.B. Fenner, J.M. Hensley, J. Vac. Sci. Technol. A 30 (2012) 061602. http://dx.doi.org/10.1116/1.4757950.
[31] J. Su, K. Deng, Z. Wang, D.-Z. Guo, 2009 IEEE Int. Freq. Control Symp. Jt. with 22nd Eur. Freq. Time forum, IEEE, 2009, pp. 1016–1018. http://dx.doi.org/10.1109/FREQ.2009.5168346.
[32] T. Baluktsian, C. Urban, T. Bublat, H. Giessen, R. Löw, T. Pfau, Opt. Lett. 35 (2010) 1950–1952. http://dx.doi.org/10.1364/OL.35.001950.
[33] P. Knapkiewicz, J. Dziuban, R. Walczak, L. Mauri, P. Dziuban, C. Gorecki, Procedia Eng. 5 (2010) 721–724. http://dx.doi.org/10.1016/j.proeng.2010.09.210.
[34] K. Tsujimoto, K. Ban, Y. Hirai, K. Sugano, T. Tsuchiya, N. Mizutani, O. Tabata, J. Micromech. Microeng. 23 (2013) 115003. http://dx.doi.org/10.1088/0960-1317/23/11/115003.

[35] A. Horsley, G.-X. Du, M. Pellaton, C. Affolderbach, G. Mileti, P. Treutlein, Phys. Rev. A 88 (2013) 063407. http://dx.doi.org/10.1103/PhysRevA.88.063407.
[36] L. Weller, R.J. Bettles, P. Siddons, C.S. Adams, I.G. Hughes, J. Phys. B: At. Mol. Opt. Phys. 44 (2011) 195006. http://dx.doi.org/10.1088/0953-4075/44/19/195006.
[37] L. Weller, T. Dalton, P. Siddons, C.S. Adams, I.G. Hughes, J. Phys. B: At. Mol. Opt. Phys. 45 (2012) 055001. http://dx.doi.org/10.1088/0953-4075/45/5/055001.
[38] L. Weller, R.J. Bettles, C.L. Vaillant, M.A. Zentile, R.M. Potvliege, C.S. Adams, I.G. Hughes, 2013, arXiv:1308.0129.
[39] L. Weller, K.S. Kleinbach, M.A. Zentile, S. Knappe, C.S. Adams, I.G. Hughes, J. Phys. B: At. Mol. Opt. Phys. 45 (2012) 215005. http://dx.doi.org/10.1088/0953-4075/45/21/215005.
[40] K.A. Whittaker, J. Keaveney, I.G. Hughes, A. Sargsyan, D. Sarkisyan, C.S. Adams, Phys. Rev. Lett. 112 (2014) 253201. http://dx.doi.org/10.1103/PhysRevLett.112.253201.
[41] S.B. Morales, E. Pangui, X. Landsheere, H. Tran, J.-M. Hartmann, Appl. Opt. 53 (2014) 4117–4122. http://dx.doi.org/10.1364/AO.53.004117.
[42] J.D. Jackson, Classical Electrodynamics, third ed., Wiley, 1999.
[43] L. Brillouin, Wave Propagation and Group Velocity, first ed., Academic Press, 1960.
[44] L.V. Hau, S.E. Harris, Z. Dutton, C.H. Behroozi, Nature 397 (1999) 594–598. http://dx.doi.org/10.1038/17561.
[45] L.J. Wang, A. Kuzmich, A. Dogariu, Nature 406 (2000) 277–279. http://dx.doi.org/10.1038/35018520.
[46] J. Keaveney, I.G. Hughes, A. Sargsyan, D. Sarkisyan, C.S. Adams, Phys. Rev. Lett. 109 (2012) 233001. http://dx.doi.org/10.1103/PhysRevLett.109.233001.
[47] D.A. Smith, I.G. Hughes, Amer. J. Phys. 72 (2004) 631. http://dx.doi.org/10.1119/1.1652039.
[48] B.E. Sherlock, I.G. Hughes, Amer. J. Phys. 77 (2009) 111. http://dx.doi.org/10.1119/1.3013197.
[49] S.J. Blundell, K.M. Blundell, Concepts in Thermal Physics, Oxford University Press, 2009, http://dx.doi.org/10.1093/acprof:oso/9780199562091.001.0001.
[50] S.R. Shin, H.-R. Noh, J. Phys. Soc. Japan 78 (2009) 084302. http://dx.doi.org/10.1143/JPSJ.78.084302.
[51] T.M. Stace, G.-W. Truong, J. Anstie, E.F. May, A.N. Luiten, Phys. Rev. A 86 (2012) 012506. http://dx.doi.org/10.1103/PhysRevA.86.012506.
[52] R.H. Dicke, Phys. Rev 89 (1953) 472–473. http://dx.doi.org/10.1103/PhysRev.89.472.
[53] V. Gerginov, A. Derevianko, C.E. Tanner, Phys. Rev. Lett. 91 (2003) 072501. http://dx.doi.org/10.1103/PhysRevLett.91.072501.
[54] B.H. Bransden, C.J. Joachain, Physics of Atoms and Molecules, second ed., Pearson Education Limited, Harlow, 2003.
[55] A. Corney, Atomic and Laser Spectroscopy, Oxford University Press, Oxford, 1977.
[56] A.R. Edmonds, Angular Momentum in Quantum Mechanics, second ed., Princeton University Press, Princeton, New Jersey, 1960.
[57] B. Schaefer, E. Collett, R. Smyth, D. Barrett, B. Fraher, Amer. J. Phys. 75 (2007) 163. http://dx.doi.org/10.1119/1.2386162.
[58] W. Yuan, W. Shen, Y. Zhang, X. Liu, Opt. Express 22 (2014) 11011–11020. http://dx.doi.org/10.1364/OE.22.011011.
[59] A. Millett-Sikking, I.G. Hughes, P. Tierney, S.L. Cornish, J. Phys. B: At. Mol. Opt. Phys. 40 (2007) 187–198. http://dx.doi.org/10.1088/0953-4075/40/1/017.
[60] C. Lee, G.Z. Iwata, E. Corsini, J.M. Higbie, S. Knappe, M.P. Ledbetter, D. Budker, Rev. Sci. Instrum. 82 (2011) 043107. http://dx.doi.org/10.1063/1.3568824.
[61] R. Clark Jones, JOSA 31 (1941) 488–493. http://dx.doi.org/10.1364/JOSA.31.000488.
[62] E. Hecht, Optics, fourth ed., Addison Wesley, 2002.
[63] J.D. Hunter, Comput. Sci. Eng. 9 (2007) 90–95. http://dx.doi.org/10.1109/MCSE.2007.55.
[64] I.G. Hughes, T.P.A. Hase, Measurements and Their Uncertainties: A Practical Guide to Modern Error Analysis, first ed., Oxford University Press, 2010.
[65] S.G. Johnson, Faddeeva function implementation, 2014. URL: http://ab-initio.mit.edu/Faddeeva.
[66] S. Russell, P. Norvig, Artificial Intelligence: A Modern Approach, second ed., Pearson Education Inc., New Jersey, 2003.
[67] S. Kirkpatrick, C.D. Gelatt Jr., M.P. Vecchi, Science 220 (1983) 671–680. http://dx.doi.org/10.1126/science.220.4598.671.
[68] N. Metropolis, A.W. Rosenbluth, M.N. Rosenbluth, A.H. Teller, E. Teller, J. Chem. Phys. 21 (1953) 1087. http://dx.doi.org/10.1063/1.1699114.
[69] M. Lundy, A. Mees, Math. Program. 34 (1986) 111–124. http://dx.doi.org/10.1007/BF01582166.
[70] C.B. Alcock, V.P. Itkin, M.K. Horrigan, Can. Metall. Q. 23 (1984) 309–313.
[71] Y. Wang, X. Zhang, D. Wang, Z. Tao, W. Zhuang, J. Chen, Opt. Express 20 (2012) 25817–25825. http://dx.doi.org/10.1364/OE.20.025817.
[72] P.M. Duarte, R.A. Hart, J.M. Hitchcock, T.A. Corcovilos, T.-L. Yang, A. Reed, R.G. Hulet, Phys. Rev. A 84 (2011) 061406. http://dx.doi.org/10.1103/PhysRevA.84.061406.
[73] D.C. McKay, D. Jervis, D.J. Fine, J.W. Simpson-Porco, G.J.A. Edge, J.H. Thywissen, Phys. Rev. A 84 (2011) 063420. http://dx.doi.org/10.1103/PhysRevA.84.063420.
[74] J. Sebastian, Ch Gross, K. Li, H.C.J. Gan, W. Li, K. Dieckmann, Phys. Rev. A 90 (2014) 033417. http://dx.doi.org/10.1103/PhysRevA.90.033417.
[75] R.I. Billmers, S.K. Gayen, M.F. Squicciarini, V.M. Contarino, W.J. Scharpf, D.M. Allocca, Opt. Lett. 20 (1995) 106–108. http://dx.doi.org/10.1364/OL.20.000106.
[76] L. Zhang, J. Tang, Opt. Commun. 152 (1998) 275–279. http://dx.doi.org/10.1016/S0030-4018(98)00199-0.
[77] A. Rudolf, T. Walther, Opt. Lett. 37 (2012) 4477–4479. http://dx.doi.org/10.1364/OL.37.004477.
[78] D. Sarkisyan, D. Bloch, A. Papoyan, M. Ducloy, Opt. Commun. 200 (2001) 201–208. http://dx.doi.org/10.1016/S0030-4018(01)01604-2.
[79] M. Bajcsy, S. Hofferberth, V. Balic, T. Peyronel, M. Hafezi, A.S. Zibrov, V. Vuletic, M.D. Luckin, Phys. Rev. Lett. 102 (2009) 203902. http://dx.doi.org/10.1103/PhysRevLett.102.203902.
[80] V. Venkataraman, K. Saha, A.L. Gaeta, Nat. Photonics 7 (2013) 138–141. http://dx.doi.org/10.1038/nphoton.2012.283.
[81] C. Perrella, P.S. Light, J.D. Anstie, F. Benabid, T.M. Stace, A.G. White, A.N. Luiten, Phys. Rev. A 88 (2013) 013819. http://dx.doi.org/10.1103/PhysRevA.88.013819.
[82] G. Epple, K.S. Kleinbach, T.G. Euser, N.Y. Joly, T. Pfau, P.St.J. Russell, R. Löw, Nature Commun. 5 (2014) 4132. http://dx.doi.org/10.1038/ncomms5132.
[83] M.R. Sprague, P.S. Michelberger, T.F.M. Champion, D.G. England, J. Nunn, X.-M. Jin, W.S. Kolthammer, A. Abdolvand, P.St.J. Russell, I. A. Walmsley, Nat. Photonics 8 (2014) 287–291. http://dx.doi.org/10.1038/nphoton.2014.45.
[84] F. Haas, J. Volz, R. Gehr, J. Reichel, J. Estève, Science 344 (2014) 180–183. http://dx.doi.org/10.1126/science.1248905.
[85] H.B.G. Casimir, D. Polder, Phys. Rev. 73 (1948) 360–372. http://dx.doi.org/10.1103/PhysRev.73.360.
[86] H. Bender, C. Stehle, C. Zimmermann, S. Slama, J. Fiedler, S. Scheel, S.Y. Buhmann, V.N. Marachevsky, Phys. Rev. X 4 (2014) 011029. http://dx.doi.org/10.1103/PhysRevX.4.011029.
[87] P. Siddons, N.C. Bell, Y. Cai, C.S. Adams, I.G. Hughes, Nat. Photonics 3 (2009) 225–229. http://dx.doi.org/10.1038/NPHOTON.2009.27.
[88] R. Löw, T. Pfau, Nat. Photonics 3 (2009) 197–199. http://dx.doi.org/10.1038/nphoton.2009.41.
[89] K.J.R. Rosman, P.D.P. Taylor, Pure Appl. Chem. 70 (1998) 217–235. http://dx.doi.org/10.1351/pac199870010217.
[90] E. Jones, T. Oliphant, P. Peterson, et al. SciPy: Open source scientific tools for Python, 2001–, URL: http://www.scipy.org/.
[91] P.J. Mohr, B.N. Taylor, D.B. Newell, Rev. Modern Phys. 84 (2012) 1527–1605. http://dx.doi.org/10.1103/RevModPhys.84.1527.
[92] P. Juncar, J. Pinard, J. Hamon, A. Chartier, Metrologia 17 (1981) 77–79. http://dx.doi.org/10.1088/0026-1394/17/3/001.
[93] U. Volz, M. Majerus, H. Liebel, A. Schmitt, H. Schmoranzer, Phys. Rev. Lett. 76 (1996) 2862–2865. http://dx.doi.org/10.1103/PhysRevLett.76.2862.
[94] S. Falke, E. Tiemann, C. Lisdat, H. Schnatz, G. Grosche, Phys. Rev. A 74 (2006) 032503. http://dx.doi.org/10.1103/PhysRevA.74.032503.
[95] H. Wang, P.L. Gould, W.C. Stwalley, J. Chem. Phys. 106 (1997) 7899. http://dx.doi.org/10.1063/1.473804.
[96] A. Banerjee, D. Das, V. Natarajan, Europhys. Lett. 65 (2004) 172. http://dx.doi.org/10.1209/epl/i2003-10069-3.
[97] U. Volz, H. Schmoranzer, Phys. Scr. T65 (1996) 48–56. http://dx.doi.org/10.1088/0031-8949/1996/T65/007.
[98] Th Udem, J. Reichert, R. Holzwarth, T.W. Hänsch, Phys. Rev. Lett. 82 (1999) 3568–3571. http://dx.doi.org/10.1103/PhysRevLett.82.3568.
[99] J.M. Amini, H. Gould, Phys. Rev. Lett. 91 (2003) 153001. http://dx.doi.org/10.1103/PhysRevLett.91.153001.
[100] J. Ye, S. Swartz, P. Jungner, J.L. Hall, Opt. Lett. 21 (1996) 1280–1282. http://dx.doi.org/10.1364/OL.21.001280.
[101] G.P. Barwood, P. Gill, W.R.C. Rowley, Appl. Phys. B 53 (1991) 142–147. http://dx.doi.org/10.1007/BF00330229.
[102] Th Udem, J. Reichert, T.W. Hänsch, M. Kourogi, Phys. Rev. A 62 (2000) 031801. http://dx.doi.org/10.1103/PhysRevA.62.031801.
[103] J.F. Sell, B.M. Patterson, T. Ehrenreich, G. Brooke, J. Scoville, R.J. Knize, Phys. Rev. A 84 (2011) 010501. http://dx.doi.org/10.1103/PhysRevA.84.010501.
[104] E. Arimondo, M. Inguscio, P. Violino, Rev. Modern Phys. 49 (1977) 31–75. http://dx.doi.org/10.1103/RevModPhys.49.31.
[105] W.A. van Wijngaarden, J. Li, Z. Phys. D 32 (1994) 67–71. http://dx.doi.org/10.1007/BF01425925.
[106] W. Yei, A. Sieradzan, M.D. Havey, Phys. Rev. A 48 (1993) 1909–1915. http://dx.doi.org/10.1103/PhysRevA.48.1909.
[107] M.P. Bradley, J.V. Porto, S. Rainville, J.K. Thompson, D.E. Pritchard, Phys. Rev. Lett. 83 (1999) 4510–4513. http://dx.doi.org/10.1103/PhysRevLett.83.4510.
[108] G. Audi, A.H. Wapstra, C. Thibault, Nucl. Phys. A 729 (2003) 337–676. http://dx.doi.org/10.1016/j.nuclphysa.2003.11.003.
[109] U.D. Rapol, A. Krishna, V. Natarajan, Eur. Phys. J. D 23 (2003) 185–188. http://dx.doi.org/10.1140/epjd/e2003-00069-9.
[110] S. Bize, Y. Sortais, M.S. Santos, C. Mandache, A. Clairon, C. Salomon, Europhys. Lett. 45 (1999) 558–564. http://dx.doi.org/10.1209/epl/i1999-00203-9.